\documentclass[aps,prl,twocolumn,showpacs,preprintnumbers,amsmath,amssymb]{revtex4}

\usepackage{graphicx}
\usepackage{dcolumn}
\usepackage{bm}

\begin{document}

\title{The Hagedorn thermostat}

\author{L. G. Moretto, K. A. Bugaev, J. B. Elliott and L. Phair}
\affiliation{Nuclear Science Division, Lawrence Berkeley National Laboratory,
Berkeley, CA 94720}

\date{\today}
\begin{abstract}
A system $\cal H$ with a Hagedorn-like mass spectrum imparts its unique temperature $T_{\cal H}$ to any other system coupled to it. An $\cal H$ system radiates particles in preexisting physical and chemical equilibrium. These particles form a saturated vapor at temperature $T_{\cal H}$. This coexistence describes a first order phase transition. An $\cal H$ system is nearly indifferent to fragmentation into smaller $\cal H$ systems. A lower mass cut-off in the spectrum does not significantly alter the general picture.
\end{abstract}

\preprint{LBNL-56898}
\maketitle

\section{Introduction}

A system~$A$ with energy $E$ and degeneracy
\begin{equation}
	\rho_A (E) \propto \exp \left( k_A E \right)
\label{therm-deg}
\end{equation}
while seemingly having a partition function of the form
\begin{equation}
	Z (T) = \int \rho_A (E) \exp\left( -{E}/{T} \right) dE
\label{therm-part}
\end{equation}
for all temperatures $T \le 1 / k_A$ in fact admits only \underline{one} temperature $T=T_A=1/k_A$ and it imparts that temperature to \underline{any} system coupled to it.

The partition function of Eq.~(\ref{therm-part}) implies that an external thermostat $B$ which, by definition has $\rho_B (E) \propto \exp \left( - k_B E \right)$, can impart its temperature $T_B =1/k_B$ to the system $A$. This is not so, as can be seen by considering the generating micro-canonical partition
\begin{eqnarray}
	P (x) & = & \rho_A (E-x) \rho_B (x) =
	\exp \left( k_A \left[ E-x \right] \right) \exp(k_B x) \nonumber \\
	& = & \exp \left[ \frac{E-x}{T_A} \right] \exp \left[ \frac{x}{T_B} \right] .
\label{two-parts}
\end{eqnarray}
The most probable partition is given by
\begin{equation}
	\frac{\partial P(x)}{\partial x} = 0 = k_A - k_B = \frac{1}{T_A} - \frac{1}{T_B} .
\label{mpp}
\end{equation}
But this is hardly possible since in general $T_A \ne T_B$: two thermostats can never be at equilibrium unless they are at the same temperature.

This preamble is motivated by the fact that the empirical hadronic mass spectra (Hagedorn spectra \cite{hagedorn-65,hagedorn-68}), the statistical bootstrap model \cite{SBM, SBM:new, SBM:width} and the MIT bag model \cite{chodos-74} have a degeneracy whose leading term is of the form of Eq.~(\ref{therm-deg}). It is the aim of this paper to explore in a pedagogical manner the implications of such a spectrum, making only passing references to the more complex physical situations occurring in particle-particle and nucleus-nucleus collisions. 

Hagedorn noted that the hadronic mass spectrum (level density) has the asymptotic ($m \rightarrow \infty$) form
\begin{equation}
	\rho_{\cal H}(m) \approx \exp \left({m}/{T_{\cal H}}\right)\,,
\label{hagedorn}
\end{equation}	
where $m$ is the mass of the hadron in question and $T_{\cal H}$ is the temperature associated with the mass spectrum \cite{hagedorn-65,hagedorn-68}. The question of the mass range over which (\ref{hagedorn}) is valid is still under discussion \cite{SBM:new,SBM:width}.

The M.I.T. bag model \cite{chodos-74} of partonic matter reproduces this behavior via a constant pressure $B$ of a ``bag'' of partonic matter \cite{kapusta-81,kapusta-82}. The pressure $p$ inside a bag at equilibrium without additional conserved quantities is
\begin{equation}
 	p=\frac{g\pi^2}{90} T^{4}_B = B\,,
\label{bag-pressure}
\end{equation}
where $g$ is the number of partonic degrees of freedom. The bag constant forces a constant temperature $T_B$ on the bag. Similarly, the enthalpy density $\epsilon$ of the bag 
\begin{equation}
 	\epsilon=\frac{H}{V}=\frac{g\pi^2}{30}T^{4}_B + B\,
\label{bag-edens}
\end{equation}
is constant.
Here $H$ is the enthalpy  and $V$ is the volume of the bag. Thus, an injection of an arbitrary amount of energy leads to an isothermal, isobaric expansion of the bag and the bag entropy $S$ is proportional to $H$:
\begin{equation}
	S = \int \frac{\delta Q}{T} = \int_{0}^{H} \frac{dH}{T} = \frac{H}{T_B}\,, 
\label{bag-eq}
\end{equation}
where $\delta Q$ is the change in heat of the bag. The bag's spectrum (level density) is then $\rho = \exp \left( S \right)$ given by Eq.~(\ref{hagedorn}) with $T_{B}=T_{\cal H}$ and $H \equiv m$.

We show here that a system $\cal H$ possessing a Hagedorn-like spectrum, characterized by an entropy of the form (\ref{bag-eq}), not only has a unique microcanonical temperature
\begin{equation}
	T_{\cal H} = \left( \frac{d S}{d E}\right)^{-1} = \left. \frac{\partial H}{\partial S} \right|_p= T_B \,,
\label{temperature}
\end{equation}
but also imparts this same temperature to any other system to which $\cal H$ is coupled. In the language of standard thermodynamics: $\cal H$ is a perfect thermostat.

The property of a perfect thermostat is well known. For instance, it is indifferent to the transfer of any portion of its energy to any parcel within itself, no matter how small. In other words, it is at the limit of phase stability and its internal fluctuations of the energy density are maximal.

\section{Harmonic Oscillator Coupled to $\cal H$}

In order to demonstrate the thermostatic behavior of a Hagedorn system, let us begin by coupling $\cal H$ to a one dimensional harmonic oscillator and use a microcanonical treatment. The unnormalized probability $P(\varepsilon)$ for finding excitation energy $\varepsilon$ in the harmonic oscillator out of the system's total energy $E$ is
\begin{eqnarray}
	P(\varepsilon) & \sim & \rho_{\cal H}(E-\varepsilon)\, \rho_{\rm osc} (\varepsilon) \nonumber \\
	&=& \exp \left( \frac{E-\varepsilon}{T_{\cal H}} \right) = \rho_{\cal H}(E) \exp \left( -\frac{\varepsilon}{T_{\cal H}} \right).
\label{sho}
\end{eqnarray}
Recall that for a one dimensional harmonic oscillator $\rho_{\rm osc}$ is a constant. The energy spectrum of the oscillator is canonical up to the upper limit ${\varepsilon}_{max} = E$ with an inverse slope (temperature) of $T_{\cal H}$ independent of $E$. The mean value of the energy of the oscillator is given by
\begin{equation}
	\overline{\varepsilon} = T_{\cal H} \left[ 1 - \frac{E / T_{\cal H}}{\exp\left( E / T_{\cal H} \right)-1} \right] .
\label{ave-ho-e}
\end{equation}
Thus in the limit that $E \rightarrow \infty$: $\overline{\varepsilon} \rightarrow T_{\cal H}$, i.e. no temperature other that $T_{\cal H}$ is admitted.

\section{An ideal vapor coupled to $\cal H$}

For a more physically relevant example, let us consider a vapor of $N\gg1$ non-interacting particles of mass $m$ coupled to $\cal H$. The microcanonical level density of the vapor with kinetic energy $\varepsilon$ is
\begin{equation}
 \rho_{\rm vapor}(\varepsilon) = \frac{V^N}{N!\left( \frac{3}{2}N \right)!} 
\left( \frac{m \varepsilon}{2\pi} \right)^{\frac{3}{2}N}\,,
\label{vapor-part}
\end{equation}
where $V$ is is the volume. The microcanonical partition of the total system is 
\begin{eqnarray}
	\rho_{\rm total}(E,\varepsilon) & = & \rho_{\cal H}(E-\varepsilon)\rho_{\rm vapor}(\varepsilon) \nonumber \\
& = & \frac{V^N}{N! \left( \frac{3}{2}N \right)!} \left( \frac{m \varepsilon}{2\pi} \right)^{\frac{3}{2}N} e^{\frac{E-mN-\varepsilon}{T_{\cal H}}} .
\label{full-part}
\end{eqnarray}
Just as with the harmonic oscillator, the distribution of the vapor is exactly canonical up to $\varepsilon_{max}=E$, if the particles are independently present, or $\varepsilon_{max}=E-mN$, if the particles are generated by $\cal H$. In either case, the temperature of the vapor is always $T_{\cal H}$.

The maximum of $\rho_{\rm total}(E,\varepsilon)$ with respect to $\varepsilon$ gives the most probable kinetic energy per particle as
\begin{equation}
	\frac{\partial \rho_{\rm total}(E,\varepsilon)}{\partial \varepsilon} = 
	\frac{3N}{2\varepsilon} - \frac{1}{T_{\cal H}} = 0 \quad \Rightarrow \quad 
	\frac{\varepsilon}{N} = \frac{3}{2}T_{\cal H}\,,
\label{max-01}
\end{equation}
provided that $E \ge mN + \frac{3}{2}N T_{\cal H}$. (For $mN < E < mN + \frac{3}{2}N T_{\cal H}$, the most probable value of the kinetic energy per particle is $\frac{\varepsilon}{N} = \frac{E}{N} - m < \frac{3}{2} T_{\cal H}$; for $E \le mN$, $\frac{\varepsilon}{N}~=~0$.~)\ \ Again $T_{\cal H}$ is the sole temperature characterizing the distribution up to the microcanonical cut-off, which may be above or below the maximum of the distribution since the form of $\rho_{\rm total}(E,\varepsilon)$ is independent of~$E$.

The maximum of $\rho_{\rm total}(E,\varepsilon)$ with respect to $N$ at fixed $V$ is given by
\begin{equation}
	\frac{\partial \ln \rho_{\rm total}(E,\varepsilon)}{\partial N} = -\frac{m}{T_{\cal H}} + \ln\left[\frac{V}{N} \left( \frac{mT_{\cal H}}{2 \pi}\right)^{\frac{3}{2}}\right]= 0,
\label{max-02}
\end{equation}
where Eq.~(\ref{max-01}) was used for $\varepsilon$. Thus the most probable particle density of the vapor is 
independent of $V$:
\begin{equation}
	\frac{N}{V} = \left( \frac{m T_{\cal H}}{2 \pi} \right)^{\frac{3}{2}} e^{-\frac{m}{T_{\cal H}}} \equiv n_{\cal H} \,.
\label{numberpp}
\end{equation}
Equation~(\ref{numberpp}) demonstrates that not only is $\cal H$ a perfect thermostat but also a perfect particle reservoir. Particles of different mass $m$ will be in chemical equilibrium with each other. At equilibrium, particles are emitted from $\cal H$ and form a saturated vapor at coexistence with $\cal H$ at temperature $T_{\cal H}$. This describes a first order phase transition (hadronic to partonic). Coexistence occurs at a single temperature fixed by the bag pressure.

These results explain the common value of: the hadronization temperatures obtained within the statistical hadronization model \cite{Becattini:1}; the inverse slopes of the transverse mass spectra of hadrons observed in high energy elementary particle collisions \cite{alexopoulos-02,Gazdzicki:04}; and the transition temperature from lattice QCD calculations for low baryonic density \cite{Lattice}. For further discussion see \cite{Bugaev:05}.

\subsection{$\cal H$ as a radiant bag}

Let us assume that $\cal H$ is a bag thick enough to absorb any given particle of the vapor striking it. Then, detailed balance requires that on average $\cal H$ radiates back the same particle. Under these conditions particles can be considered to be effectively emitted from the surface of $\cal H$. Thus the relevant fluxes do not depend in any way upon the inner structure of $\cal H$. 

In fact, the results given in equations (\ref{max-01}) and (\ref{numberpp}) show that the saturated vapor concentration depends only upon $m$ and $T_{\cal H}$ as long as $\cal H$ is present. A decrease in the volume $V$ does not increase the vapor concentration, but induces a condensation of the corresponding amount of energy out of the vapor and into $\cal H$. An increase in $V$ keeps the vapor concentration constant via evaporation of the corresponding amount of energy out of $\cal H$ and into the vapor. This is reminiscent of liquid-vapor equilibrium at fixed temperature, except that here coexistence occurs at a single temperature $T_{\cal H}$, rather than over a range of temperatures as in ordinary fluids.

The bag wall is Janus faced: one side faces the partonic world, and, aside from conserved charges, radiates a partonic black body radiation responsible for balancing the bag pressure; the other side faces the hadronic world and radiates a hadronic black body radiation, mostly pions. Both sides of the bag wall are at the temperature $T_{\cal H}$. It is tempting to attribute most, if not all, of the hadronic and partonic properties to the wall itself, possibly even the capability to enforce conservation laws globally (quantum number conductivity). Despite the fact that this wall is an insurmountable horizon with hadronic measurements such as bag size and total radiance we can infer some properties of the partonic world, e.g. the number of degrees of freedom \cite{alexopoulos-02}.

We can estimate an upper limit for the emission time using the outward energy flux of particles radiated from the bag. At equilibrium the in-going and out-going fluxes must be the same, thus the outward flux of particles 
in the nonrelativistic approximation using Eq.~(\ref{numberpp}) is
\begin{equation}
	\varphi_{n_{\cal H}} \simeq \frac{n_{\cal H}}{4} \left( \frac{m}{m+2T_{\cal H}}\right) \sqrt{8\frac{T_{\cal H}}{\pi m}}\,. 
\label{part-flux}
\end{equation}
Using the technique developed in \cite{Bugaev:96, Bugaev:99}, one finds the energy flux $\varphi_{E_{\cal H}}$ and momentum flux 
$p_{\rm rad}$ as
\begin{equation}
	\varphi_{E_{\cal H}} \simeq \left( m +2 T_{\cal H}\right) \varphi_{n_{\cal H}} \,, \quad 
	p_{\rm rad} = \frac{1}{2} n_{\cal H} T_{\cal H}\,.
\label{enrgy-flux}
\end{equation}
The pressure $p_{\rm rad}$ exerted on the bag by its radiation can be compared to the intrinsic bag pressure in Eq.~(\ref{bag-pressure}): for pions $p_{\rm rad} \sim 0.02 B$. The time $\tau$ for the bag to dissolve into its own radiation is approximately
\begin{equation}
\tau \simeq \frac{3 \pi \exp 
\left( \frac{m}{T_{\cal H}} \right) E_0}{g_m \left( m^2 T_{\cal H}^{2} \right ) R_{0}^{2}}\,,
\label{bag-time}
\end{equation}
where $g_m$ is the degeneracy of mass $m$ particles, $R_0$ is the initial bag radius and $E_0$ is the initial bag total energy.

The fluxes written in Eqs.~(\ref{part-flux}) and (\ref{enrgy-flux}) (particle or energy per unit surface area) 
are integrated over an assumed spherical bag to give the result in Eq.~(\ref{bag-time}). However, because 
of the lack of surface tension, the bag's maximum entropy corresponds to either an elongated (cylinder) or 
a flattened shape (disc). Thus, Eq.~(\ref{bag-time}) should be interpreted as an upper limit. More detailed studies of hadron emission from bags concerning hydrodynamic shock waves and freeze out shocks can be found elsewhere \cite{Bugaev:96, Bugaev:99,studies}.

The decoupling between the vapor concentration and $m$ and $T_{\cal H}$ occurs when $\cal H$ has completely evaporated (i.e. when $E-Nm-\frac{3}{2}NT_{\cal H}=0$) at a volume of 
\begin{equation}
	V_{d} \simeq \frac{1}{n_{\cal H}} \frac{E}{m+\frac{3}{2}T_{\cal H}} .
\label{v-decoup}
\end{equation}
The disappearance of $\cal H$ allows the vapor concentration to decrease inversely proportionally to $V$ as
\begin{equation}
	\frac{N}{V} = \frac{n_{\cal H}V_{d}}{V} .
\label{conc-dec}
\end{equation}

\begin{figure}[ht]
\includegraphics[width=8.7cm,height=7.0cm]{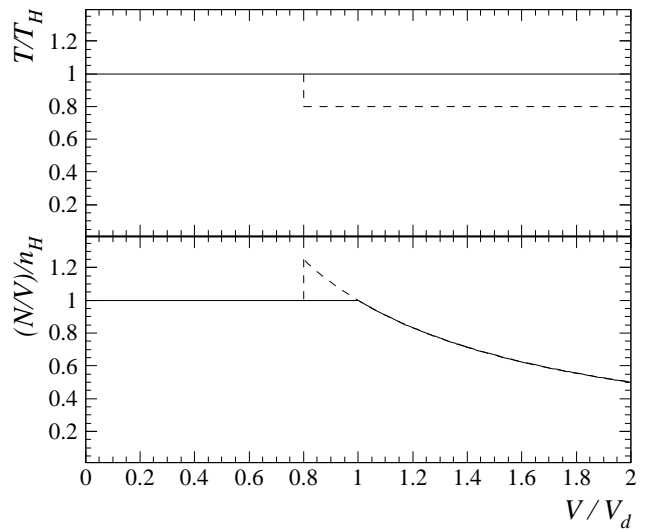}
\caption{Typical behavior of the entire system's temperature $T$ and concentration $N/V$ as the function of
the system's volume $V$ in the absence of restrictions (solid curve) and for a finite cut off at $m_0$ of the Hagedorn spectrum (dashed curve).}
\label{schematic}
\end{figure}

The temperature, however, remains fixed at $T_{\cal H}$ because of conservation of energy and particle number above $V_{d}$.\ \ Solid curves in Fig.~\ref{schematic} show this schematically.

The discussion above assumes that the Hagedorn spectrum extends down to $m=0$. However, experimentally there appears to be a lower cut off of the spectrum at $m_0$. This modifies the above results as follows.

For energies $E - mN - \varepsilon \gg m_0$ and $V<V_d$ the above results hold as written. However, if we increase the volume well beyond $V_d$ at which the Hagedorn spectrum is truncated at $m_0$, the situation is slightly different. $\cal H$ evaporates until its mass is $m_0$. If the entire mass of $\cal H$ is fully transformed into vapor particles as the volume is increased further, then the excess particles temporarily increase the concentration and permanently decrease the temperature. As the volume increases further, the concentration changes inversely proportional to $V$
\begin{equation}
	\frac{N}{V} = \frac{n_{\cal H} V_{d}+\frac{m_0}{m}}{V} ,
\label{conc-dec2}
\end{equation}
\mbox{while the temperature remains constant at}
\begin{equation}
	T = \frac{n_{\cal H} V_d}{n_{\cal H} V_{d}+\frac{m_0}{m}} T_{\cal H} .
\label{conc-temp}
\end{equation}
Dashed curves in Fig.~\ref{schematic} show this schematically.

\section{Fragmentation of $\cal H$}

A question of interest is the stability of $\cal H$ against fragmentation. If the translational degrees of freedom are neglected, $\cal H$ is indifferent to fragmentation into an arbitrary number of particles of arbitrary mass (within the constraints of mass/energy conservation).

Let us now consider the case in which the mass of the vapor particle $m$ is allowed to be free. The system's level density $\rho_{\rm total}(E,\varepsilon)$ is still given by Eq.~(\ref{full-part}). Using Eqs.~(\ref{max-01}) and~(\ref{numberpp}), one finds the most probable value of the system's level density as $\rho_{\rm total}^*(E,\varepsilon) \approx \exp\left[ S^* \right]$, where the entropy is $S^* = E/T_{\cal H} + N $. Differentiating $\rho_{\rm total}^*(E,\varepsilon)$ with respect to $m$ and applying Eq.~(\ref{numberpp}) gives
\begin{equation}
\hspace{-.004cm}\frac{\partial \ln \rho_{\rm total}^*(E,\varepsilon)}{\partial m} = N  \left[ \frac{3}{2 m}- 
\frac{1}{T_{\cal H} } \right] \,\,\, \Rightarrow \,\,\,  m = \frac{3}{2}T_{\cal H}\,,
\label{masspp}
\end{equation}
i.e. the last equality provides the maximum of level density for $N \neq 0$. Since all the intrinsic statistical weights in $\rho_{\rm total}^*(E,\varepsilon)$ are factored into a single $\cal H$ the system breaks into fragments with $m=\frac{3}{2}T_{\cal H}$ except for one whose mass is determined by mass/energy conservation.

Substituting the most probable value of $\varepsilon$ and $m$ into the most probable value of $N$ one obtains the vapor concentration
\begin{equation}
	\frac{N}{V} = \left( \frac{3}{4 \pi e} \right)^{\frac{3}{2}} T_{\cal H}^{3}\,.
\label{concpp}
\end{equation}
The density of the vapor of nonrelativistic particles acquires the form typical of the ultrarelativistic limit.

If the value of mass given by Eq.~(\ref{masspp}) does not exist, then the most probable
value of level density $\rho_{\rm total}^*(E,\varepsilon) $ corresponds to the mass $m^*$ which is nearest to 
$\frac{3}{2}T_{\cal H}$ 
and $N (m^*) $ given by Eq. (\ref{numberpp}). 
In terms of hadron spectroscopy the value of $m^*$ that maximizes  
the level density $\rho_{\rm total}^*(E,\varepsilon) $
is the pion mass. 

If $\cal H$ is required to fragment totally into a number of equal fragments all endowed with their translational degrees of freedom, then
\begin{eqnarray}
\rho_T & = &
 \frac{ {e^{\frac{E - \varepsilon}{ T_{\cal H} } } }
 V^N}{N! \left( \frac{3}{2}N\right)!} 
 \left[ \frac{m \varepsilon}{2 \pi} \right]^{\frac{3}{2}N} \nonumber \\
 & = &
 \frac{ {e^{\frac{E}{ T_{\cal H} } } } ~ V^N}{N! } \left[ \frac{m T_{\cal H}}{2 \pi} \right]^{\frac{3}{2}N} ,
\label{equation}
\end{eqnarray}
where in the last step we substituted the most probable value of the kinetic energy (\ref{max-01}) 
and used the Stirling formula for $ \left( \frac{3}{2}N\right)!$. From 
Eq.~(\ref{equation}) it is seen that all the Hagedorn factors collapse into a single one with the 
$m$-independent argument $E$. Maximization of (\ref{equation}) with respect to $m$ leads to
\begin{equation}
	\frac{\partial \ln \rho_T}{\partial m} = \frac{3N}{2m} = 0\,,
\label{max-eq}
\end{equation}
which is consistent with $N=1$ and $m=E$, namely a single Hagedorn particle with all the available mass. 
This result justifies the assumption of the canonical formulation of the statistical hadronization model that smaller clusters appear from a single large cluster \cite{Becattini:Can}.

This again illustrates the indifference of $\cal H$ toward fragmentation. Of course Eq.~(\ref{max-01}) gives directly the mass distribution of the Hagedorn fragments under the two conditions discussed above.

\section{Conclusions}

A system $\cal H$, with a Hagedorn-like mass spectrum, is a perfect thermostat and a perfect particle reservoir. Consequently, any system coupled to $\cal H$ can have only the temperature of $\cal H$: $T_{\cal H}$. This behavior may explain the common value of: the hadronization temperatures obtained within statistical models; the transition temperature from lattice QCD calculations for low baryonic density; and the inverse slopes of the transverse mass spectra of hadrons (temperature) observed in high energy elementary particle collisions and high energy nucleus-nucleus collisions  (for details see \cite{Bugaev:05}). The common temperature of the experimental spectra may indicatde the observed particles originate from an $\cal H$-like system.

The hadronic side of $\cal H$ radiates particles in preexisting physical and chemical equilibrium just as a black body radiates photons in physical and chemical equilibirum (compare to Ref. \cite{Greiner:04}). Particles emitted from $\cal H$ form a saturated vapor that coexists with $\cal H$. This coexistence describes a first order phase transition (hadronic to partonic) and occurs at a single temperature fixed by the bag pressure. An $\cal H$ system is nearly indifferent to fragmentation into smaller $\cal H$ systems. A lower cut-off in the mass spectrum does not alter our results \cite{Bugaev:05}.

\end{document}